\documentclass[sort&compress]{aipproc}
\usepackage{epsfig}
\usepackage{axodraw}
\layoutstyle{6x9}

\begin{document}

\title{The pion form factor from first principles}

\author{J. van der Heide}{
address={National Institute for Nuclear Physics and High-Energy Physics (NIKHEF), 1009 DB Amsterdam, The Netherlands},
email={r86@nikhef.nl}
}

\begin{abstract}
We calculate the electromagnetic form factor of the pion in quenched lattice
QCD. The non-perturbatively improved Sheikoleslami-Wohlert
lattice action is used together with the ${\mathcal O}(a)$ improved
current. We calculate form factor for pion masses down
to $m_\pi = 360\; MeV$.
We compare the mean square radius for the pion extracted from our
form factors to the value obtained from the 'Bethe Salpeter
amplitude'. Using (quenched) chiral perturbation theory, we extrapolate
our results towards the physical pion mass.
\end{abstract}

\maketitle

\section{Introduction}
The pion, being the lightest and simplest particle in the hadronic spectrum has been studied intensively in the past. Using a variety of effective and phenemenological models, the properties of the pion have been desribed with varying succes. However, these models make assumptions, for example, confinement is put in by hand in contrast to being the result of the underlying dynamics. Lattice QCD (LQCD) doesn't have this drawback since it is solves QCD directly.

Using LQCD, global properties of the pion such as the mass and the decay width have been calculated to satisfying accuracy. The form factor, which directly reflects the internal structure, is clearly an important challenge. The first lattice results were obtained by Martinelli and Sachrajda \cite{Martinelli}. It was followed by a more detailed study by Draper {\it et al.} \cite{Draper}, 
who showed that the form factor could be described by a simple monopole form as suggested 
by vector meson dominance \cite{Williams}. We extend \cite{us} these studies by adopting improved lattice techniques \cite{Luscher,Sheikholeslami,Sachrajda,Luscher2,Sommer}, which means that we include extra operators in order to systematically eliminate all the $\mathcal{O}(a)$ discretisation errors. 

Some aspects of the pion structure have been obtained \cite{Chu,Hecht,Gupta,Lacock,Schmidt} using 'the Bethe-Salpeter method'. We also use this approach and compare its predictions to the results of our direct calculation of the pion form factor.

Finally we study a chiral extrapolation to reach the physical limit.

\section{The method}
To extract the form factor we calculate the two- and three point functions of the pion, analogously to \cite{Draper}. The two point function is given by \begin{equation}
G_2 (t, {\bf p}) = \sum_{\bf x} \left<\phi({t, \bf x})\; \phi ^{\dagger}
(0,{\bf 0})\right> \; e^{i\; {\bf p}\cdot {\bf x}} \; ,
\end{equation}
where $\phi^{\dagger}$ is an operator creating a state with the 
quantum numbers of the pion.
By varying the interquark distance at the sink, $t_f$, we can improve the overlap with the physical pion and obtain information on the 'Bethe-Salpeter amplitudes'.
The three point function is calculated as
\begin{equation}
G_3 (t_f, t; {\bf p}_f , {\bf  p}_i) = \sum_{{\bf x}_f, {\bf x}} \;
\left<\phi (x_f) \; j_4 (x) \; \phi^{\dagger} (0)\right>  e^{-i\; {\bf
p}_f \cdot ({\bf x}_f - {\bf x}) \; - i \; {\bf p}_i \cdot {\bf x}} 
\end{equation}
with $j_4$ the fourth component of the current, inserted at time $t$.
Since the local current 
\begin{equation}
j_{\mu}^{\>L} (x) = {\bar \psi (x) }\; \gamma_\mu \; \psi (x),
\end{equation}
is not conserved on the lattice, one can construct the Noether current belonging to our action
\begin{equation}
j_{\mu}^{\>C}=\kappa\left({\bar\psi(x)}(1-\gamma_\mu)U_\mu(x)\psi(x+{\hat
\mu})-{\bar\psi(x+{\hat\mu})}(1+\gamma_\mu)U^{\dagger}_\mu (x)\psi (x) 
\right).
\end{equation}
This current however, still has $\mathcal{O}(a)$ corrections for $Q^2\; > \; 0$. A conserved and improved current can be constructed \cite{Sachrajda,Luscher2,Sommer}
\begin{equation}
j_{\mu}^{\>I} = Z_{V} \, \{ j_{\mu}^{\>L} (x)  + a \> c_V \> \partial _\nu \, 
T_{\mu \nu} \} \;, 
\end{equation} 
with 
\begin{eqnarray} 
T_{\mu \nu} & = & {\bar \psi}(x)\; i \; \sigma_{\mu \nu}  \; \psi (x) 
\ , \\
Z_V & = & Z_V^{\>0} \: (1+ a\,b_V\,m_q) \; .\nonumber
\end{eqnarray}
The bare-quark mass is defined as $m_q=\frac{1}{2a}\left(1/\kappa-1/\kappa_{c}\right)$,
where $\kappa_{c}$ is the kappa value in the chiral limit and $a$ denotes the lattice spacing. For our simulation, $\kappa_{c}=0.13525$ \cite{Bowler}. 
Comparison of the currents will give us information on the importance of improvement.

\section{Details of the calculation}
On a $24^3x32$ lattice, we generated $\mathcal{O}(100)$ quenched gluon configurations at $\beta=6.0$.
Subsequently, we calculated non-perturbatively improved ($c_{SW}=1.796$ \cite{Luscher}) quark 
propagators for 5 different values of the hopping parameter, $\kappa$. These propagators were then combined to two- and three point functions for the pion with masses ranging from 360 to 970 MeV.\footnote{The lattice spacing $a = 0.105\; {\rm fm}$ is taken from \cite{Edwards}.} 
For the improved current, we use the parameters $Z_V^{\>0}$, $b_V$ and 
$c_V$ as determined by Bhattacharya {\it et al.} \cite{Bhattacharya}.
To extract the significant parameters from our numerical data, we use the following parametrisations. For the two point function, we have
\begin{equation}
G_2 (t,{\bf p}) = \sum_{n=0}^1{\sqrt {Z_R^n({\bf p})\: Z_0^n({\bf p})}}
\left\{e^{-E^n_{\bf p}t}+e^{-E^n_{\bf p}(N_{\tau}-t)}\right\}\; ,
\end{equation}
including the contribution of the ground state (n=0) and a first excited
one (n=1). The $Z^n_R$ denote the matrix elements,
\begin{equation}
Z^n_R({\bf p}) \equiv |\left<\Omega | \phi_R | n, {\bf p} \right>|^2 \;,
\end{equation}
and are related to the 'Bethe-Salpeter amplitude', $\Phi(R)=\sqrt {Z^\pi_R ({\bf 0}) \, / \, Z^\pi_0 ({\bf 0})}$.

$E^0_{\bf p}$ and $E^1_{\bf p}$ are the energies of ground and excited
state, respectively; $R$ denotes the interquark distance.
The three-point function is parametrised as
\begin{eqnarray}
G_3(t_f,t;{\bf p}_f,{\bf  p}_i) = F(Q^2) \; \sqrt {Z_R^0 ({\bf p}_f)\;
Z_0^0 ({\bf p}_i) } e^{- E^0_{{\bf p}_f }\; (t_f - t)\,  - E^0_{{\bf
p}_i}\; t } \nonumber\\
+\left\{\sqrt {Z_R^1 ({\bf p}_f) Z_0^0 ({\bf p}_i)}\left<1,{\bf p}_f|
j_\mu (0) |0,{\bf p}_f \right>e^{-\, E^1_{{\bf p}_f} \, (t_f - t)  -\,
E^0_{{\bf p}_i} \, t} \;+ (1 \leftrightarrow 0) \right\}\;.
\end{eqnarray}
The production of pion pairs, 'wrap around effects' being due to the propagation of states beyond $t_f$ and an elastic contribution from the excited state were ignored in our parametrisation after being estimated negligable. 

All parameters in the two- and three-point functions - energies $E$,
$Z$-factors and the form factor $F(Q^2)$ - were fit simultaneously to
the data per configuration. For the three-point function, we let the current insertion time $t$ vary from $0$ to $t_f$.
The value for the parameters and their error in these simultaneous fits
are  obtained through a single elimination jackknife procedure.

\section{Results}
As a byproduct of our simulations we also obtain pion masses for the 5 different $\kappa$-values. They agree with the literature. We also checked the energie-momentum relation and up to the energies involved we found that a continuum relation provides the best description. These non-trivial tests indicate that our simulations are done correctly.

\begin{figure}
\includegraphics[height=80mm]{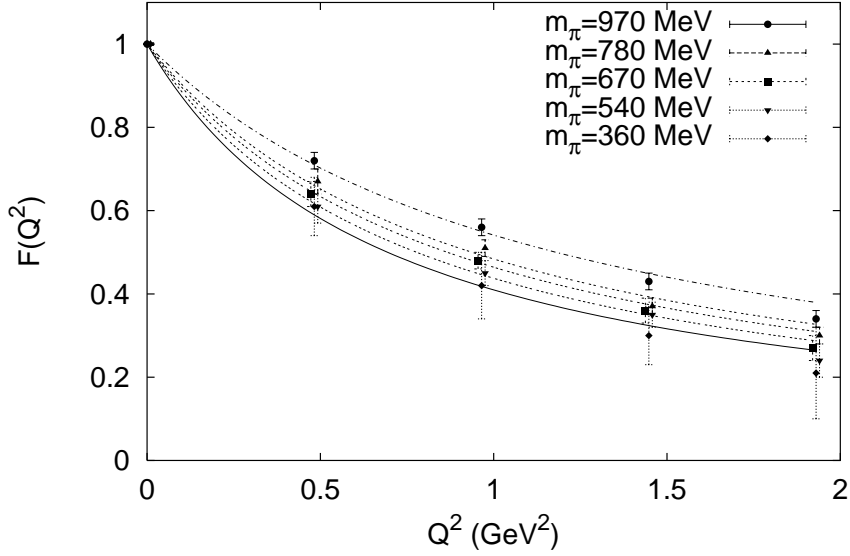}
\caption{{\em Form factors as a function of $Q^2$ for the five pion 
masses. Curves: monopole fits to the data.}}
\label{fig:ff_VMD}
\end{figure}

Using the different currents, we extract the form factor for our five $\kappa$-values. Comparison of the results for the Noether current and the improved current yields that the effect of improvement can be as large as 25$\%$ for the highest momenta considered.
The improved results are shown in Fig.\ref{fig:ff_VMD}.
The high accuracy for the datapoint at $Q^2=0$ is due to the fact that we satisfy the Ward-Takahashi identity to 1 ppm.

\begin{figure}[!h]
\includegraphics[height=80mm]{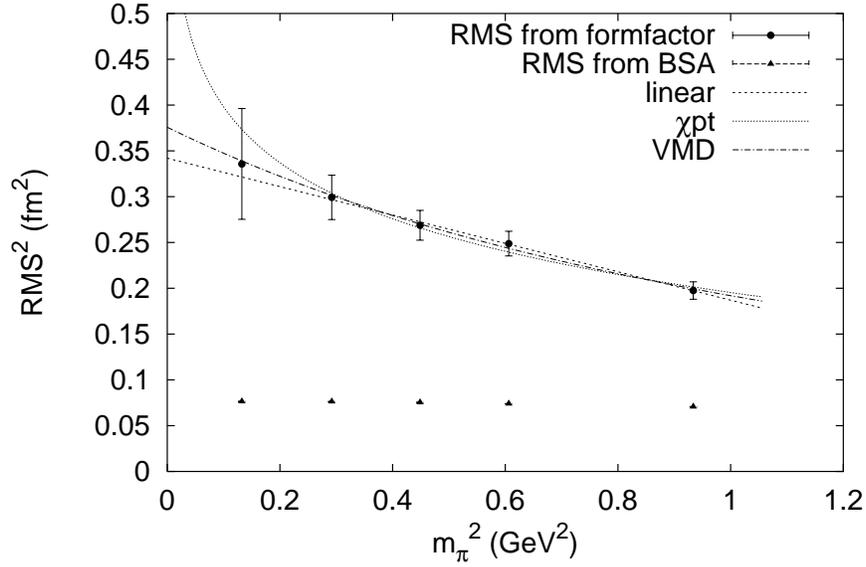}
\caption{\label{fig:RMS}
{\em Radius of the pion as obtained from Bethe-Salpeter amplitudes,
   Eq.~\ref{eq:RMS_BSA}, and from the form factor, Eq.~\ref{eq:RMS_FF}.
   Also shown are three different parametrizations of
   $\left< r^2 \right>$ (see text).}
   }
\end{figure}

As in the previous study \cite{Draper} of the pion form factor we compare our results to a monopole form factor
\begin{equation}
F(Q^2) = \{ 1 + \frac{Q^2}{M_\rho^2}\}^{-1} \;,
\label{eq:VMD}
\end{equation}
which is suggested by the vector meson dominance ansatz. 
Fitting our data to this model, we extract a vector meson mass which is within 5$\%$ of the corresponding rho mass on the lattice \cite{Bowler}.
From the behaviour of the form factor at low $Q^2$, we can extract the mean-square charge radius of the pion,
\begin{equation}
\left. \frac{dF(Q^2)}{dQ^2}\right|_{Q^2=0}=-\frac{1}{6}\left<
r^2\right>_{FF} = -\frac{1}{M_{V}^2}
\label{eq:RMS_FF}
\end{equation}
where in the last step we assume Eq.~\ref{eq:VMD} and use the fitted parameter $M_V$.
The results are shown in Fig.~\ref{fig:RMS} as a function of the pion mass. 
Previously, the 'Bethe-Salpeter-amplitude' $\Phi(R)$ has been used to obtain 
estimates of the charge radius,
\begin{equation}
\left<r^2\right>_{BS} := \frac{1}{4}
\frac{
\int d^3 \vec r \; \vec r^{\,2} \; \Phi^2(|\vec r|)
}{
\int d^3 \vec r \; \Phi^2(|\vec r|)
}\; .
\label{eq:RMS_BSA}
\end{equation}
The results based on this procedure are also shown in Fig.~\ref{fig:RMS}.
As can be seen these values are much lower than the actual values obtained from the form factor. 
Moreover, the Bethe-Salpeter results are almost mass independent, in accordance with
the observations of Refs.~\cite{Chu,Hecht,Gupta,Schmidt}. This is a known\cite{Gupta} 
deficit of the approach, which Fig.~\ref{fig:RMS} makes quantitative.
We extrapolate our results obtained with Eq.~\ref{eq:RMS_FF} using three different parametrisations. 
First, we try chiral perturbation theorie ($\chi$pt). 
At one-loop order the prediction for the radius \cite{GL} is
\begin{equation}
\left< r^2\right>_{\chi {\rm PT}} ^{\rm one-loop}
= c_1 + c_2 \log m_\pi^2
\label{eq:cPT}
\end{equation}
In quenched $\chi$pt the rms is constant to this order. There are however indications that a mass 
dependence appears at higher order \cite{Colangelo1}. For our masses we restrict ourselves to a term linear in $m_{\pi}^2$ \cite{Colangelo2}. Lastly, we have also used the VMD prescription and assume a linear dependence of $M_V$ on $m_{\pi}^2$. These three expectations are also plotted in Fig.~\ref{fig:RMS}.
\newpage

\section*{Acknowledgements}
This work has been done in collaboration with Justus Koch and Edwin Laermann.

\end{document}